\documentclass[11pt]{article}

\usepackage[utf8]{inputenc}
\usepackage[T1]{fontenc}
\usepackage{lmodern}
\usepackage[margin=1in]{geometry}
\usepackage{microtype}

\usepackage{amsmath, amssymb, mathtools}

\usepackage{graphicx}
\usepackage{pifont}
\usepackage{pgfplots}
\pgfplotsset{compat=1.18}
\usepackage{booktabs}
\usepackage{siunitx}

\usepackage[backend=biber, style=numeric-comp, sorting=none]{biblatex}
\usepackage[colorlinks=true, linkcolor=blue, citecolor=blue, urlcolor=blue]{hyperref}
\usepackage[capitalize]{cleveref}

\newcommand{\bibby}{Bibby~AI}

\title{Bibby AI: An Editor-Native Agentic Platform for\\Academic Research, Writing, and Publishing}
\author{Nilesh Jain\thanks{Founder, Bibby AI (\url{https://trybibby.com}). Contact: \texttt{nilesh@trybibby.com}}}
\date{\today}

\begin{document}

\maketitle

\begin{abstract}
Academic output is produced across a fragmented toolchain: literature discovery in one application, reference management in another, writing in a LaTeX editor, formatting against venue templates by hand, and submission through yet another portal.
Each boundary between tools forces a context switch, a format conversion, or a manual copy--paste step, and the cumulative cost dominates the time researchers spend on activities that are not research.
We present \bibby{}, an editor-native platform that collapses this toolchain into a single Research--Write--Publish pipeline built around a cloud LaTeX editor.
Unlike assistants that attach to an existing editor through a browser extension, \bibby{} owns the full document state, compilation pipeline, and revision history, which allows its agents to perform retrieval-grounded citation insertion, structural edits, and template-compliant reformatting as first-class, verifiable operations rather than text suggestions.
The platform integrates (i) ingestion pipelines that convert PDF, DOCX, and handwritten mathematics into clean LaTeX; (ii) a retrieval layer over scholarly metadata enriched with patent-to-paper citation signals derived from USPTO PatentsView and the Marx--Fuegi citation corpus, surfacing the translational impact of candidate references; and (iii) task-scoped agents for literature triage, drafting, revision, and venue formatting that operate directly on the document's abstract syntax representation.
\bibby{} is deployed in production and serves 5{,}000+ active researchers across 50+ subscribing universities.
We describe the architecture, the design decisions that editor-nativeness makes possible, and the workflow-level time-savings framework we use to evaluate the platform against fragmented baselines.
\end{abstract}

\section{Introduction}\label{sec:intro}

The median research paper is written against a stack of disconnected tools: a search engine or scholarly index for discovery, a reference manager for bibliography curation, a LaTeX editor for composition, venue style files wrestled into compliance by hand, and a submission system at the end.
Large language models (LLMs) have been bolted onto individual stages of this pipeline---grammar polishing, chat-based question answering, citation lookup---but the assistants remain \emph{external} to the environment where the document actually lives.
Prior work has shown that this separation prevents deep interaction with document state, structure, and revision history, and has attempted to bridge it with browser extensions that synchronize bidirectionally with editors such as Overleaf~\cite{hou2025paperdebugger}.

We argue for the stronger position: the assistant should not be bridged \emph{into} the editor; the editor, the compiler, the reference graph, and the agents should be a single system.
\bibby{} (\url{https://trybibby.com}) is built on this premise as a standalone, full replacement for cloud LaTeX editors such as Overleaf---not an extension, plugin, or overlay on any host platform.
Its core is a cloud LaTeX editor with server-side compilation, and every assistant capability is implemented as an operation on the platform's own document model rather than as text injected into a third-party interface.
This eliminates the hardest engineering problems that plague plugin architectures---editor synchronization, patch conflicts, and state security across origin boundaries~\cite{hou2025paperdebugger}---by construction, and it unlocks capabilities that are not expressible from outside the editor at all: compilation-verified edits, bibliography-aware refactoring, and one-click retargeting of a manuscript to a different venue template.

\paragraph{Contributions.} This paper makes four contributions:
\begin{enumerate}
  \item \textbf{An editor-native architecture} for agentic academic writing in which agents operate on the platform's document and compilation state directly, with every structural edit validated by the compiler before it is surfaced to the user (\cref{sec:architecture}).
  \item \textbf{Ingestion pipelines} that convert the formats researchers actually start from---PDF, DOCX, and handwritten mathematics---into clean, compilable LaTeX, removing the largest single onboarding cost of LaTeX-based workflows (\cref{sec:ingestion}).
  \item \textbf{A translational-impact retrieval layer} that enriches standard scholarly metadata with patent-to-paper citation signals from USPTO PatentsView~\cite{patentsview} and the Marx--Fuegi front-page citation corpus~\cite{marx2020reliance}, letting authors see which candidate references have demonstrated downstream technological impact---a signal no incumbent writing platform surfaces (\cref{sec:retrieval}).
  \item \textbf{A deployment report and evaluation framework}: \bibby{} serves 5{,}000+ active researchers and 50+ subscribing universities in production, and we define a workflow-level accounting of researcher time savings that we use for ongoing measurement (\cref{sec:deployment,sec:evaluation}).
\end{enumerate}

\section{Related Work}\label{sec:related}

\paragraph{Cloud LaTeX editors.}
Overleaf is the incumbent collaborative LaTeX platform and has added an AI Assist add-on offering writing suggestions and TeX error explanations.
These are assist-level features: they annotate the author's work but do not execute multi-step workflows, ingest external formats into compiled projects, or ground citation choice in retrieval.
\bibby{} competes at the platform level---editor, compilation, collaboration---while differentiating on agentic workflow execution, ingestion pipelines, and impact-aware retrieval (\cref{tab:comparison}).

\paragraph{In-editor writing assistants.}
PaperDebugger~\cite{hou2025paperdebugger} is the closest system in spirit: a Chrome extension that attaches a multi-agent assistant to Overleaf, with a Kubernetes-orchestrated backend and an MCP toolchain for literature search and document scoring.
Its authors identify bidirectional editor synchronization, fine-grained patching, and secure state management as the central technical obstacles of the plugin approach.
\bibby{} sidesteps this class of problem: because the editor is ours, there is no synchronization boundary, no reverse-engineered document state, and no dependency on a host platform's goodwill or API stability.
XtraGPT~\cite{chen2025xtragpt} studies controllable, context-aware paper revision as a model-level problem; \bibby{} consumes such revision capabilities as one agent among several, grounded in full-document context that the platform natively holds.

\paragraph{Scholarly retrieval infrastructure.}
Open scholarly indices such as Semantic Scholar~\cite{kinney2023semanticscholar} and OpenAlex~\cite{priem2022openalex} provide the citation graph and metadata backbone on which modern discovery tools are built.
\bibby{} builds its retrieval layer on these corpora and differentiates by joining them against patent-side evidence: PatentsView's disambiguated USPTO data~\cite{patentsview} and the Marx--Fuegi corpus of front-page patent citations to science~\cite{marx2020reliance}, which established at scale that patent-to-paper citations are a meaningful, measurable signal of science's technological reliance.

\paragraph{Document conversion.}
Format conversion into LaTeX is typically handled by standalone utilities (e.g., Pandoc for DOCX, OCR-based math recognition systems for handwriting).
These tools are disconnected from the destination editor, so their output errors surface only at first compile, in a different application.
\bibby{} integrates conversion into the platform so that output is compiled, validated, and opened as a live project in one step.

\section{System Architecture}\label{sec:architecture}

\begin{table}[t]
  \centering
  \caption{Capability comparison. ``Extension-based'' denotes plugin assistants attached to a host editor (e.g., PaperDebugger on Overleaf~\cite{hou2025paperdebugger}). \ding{51}~= native; $\sim$~= partial or add-on; \ding{55}~= absent.}
  \label{tab:comparison}
  \begin{tabular}{lccc}
    \toprule
    Capability & Overleaf & Extension-based & \bibby{} \\
    \midrule
    Cloud LaTeX editing \& compilation & \ding{51} & host-dependent & \ding{51} \\
    Writing suggestions / error help & $\sim$ (add-on) & \ding{51} & \ding{51} \\
    Compile-verified agentic edits & \ding{55} & \ding{55} & \ding{51} \\
    Multi-step workflow agents & \ding{55} & \ding{51} & \ding{51} \\
    PDF/DOCX/handwriting $\to$ project ingestion & \ding{55} & \ding{55} & \ding{51} \\
    In-editor retrieval $\to$ BibTeX insertion & \ding{55} & $\sim$ & \ding{51} \\
    Patent-to-paper impact signal & \ding{55} & \ding{55} & \ding{51} \\
    One-click venue retargeting & \ding{55} & \ding{55} & \ding{51} \\
    No host-platform dependency & \ding{51} & \ding{55} & \ding{51} \\
    \bottomrule
  \end{tabular}
\end{table}

\bibby{} is organized as four layers sharing one source of truth: the project store, which holds the LaTeX source tree, compiled artifacts, bibliography, and revision history for every project.

\subsection{Editor and Compilation Core}\label{sec:editor}
The editor is a browser-based LaTeX environment backed by server-side \texttt{pdflatex} compilation in isolated containers, with per-compile resource limits (CPU, memory, and file-descriptor caps) tuned from production incident experience.

\begin{figure}[t]
  \centering
  \includegraphics[width=\linewidth]{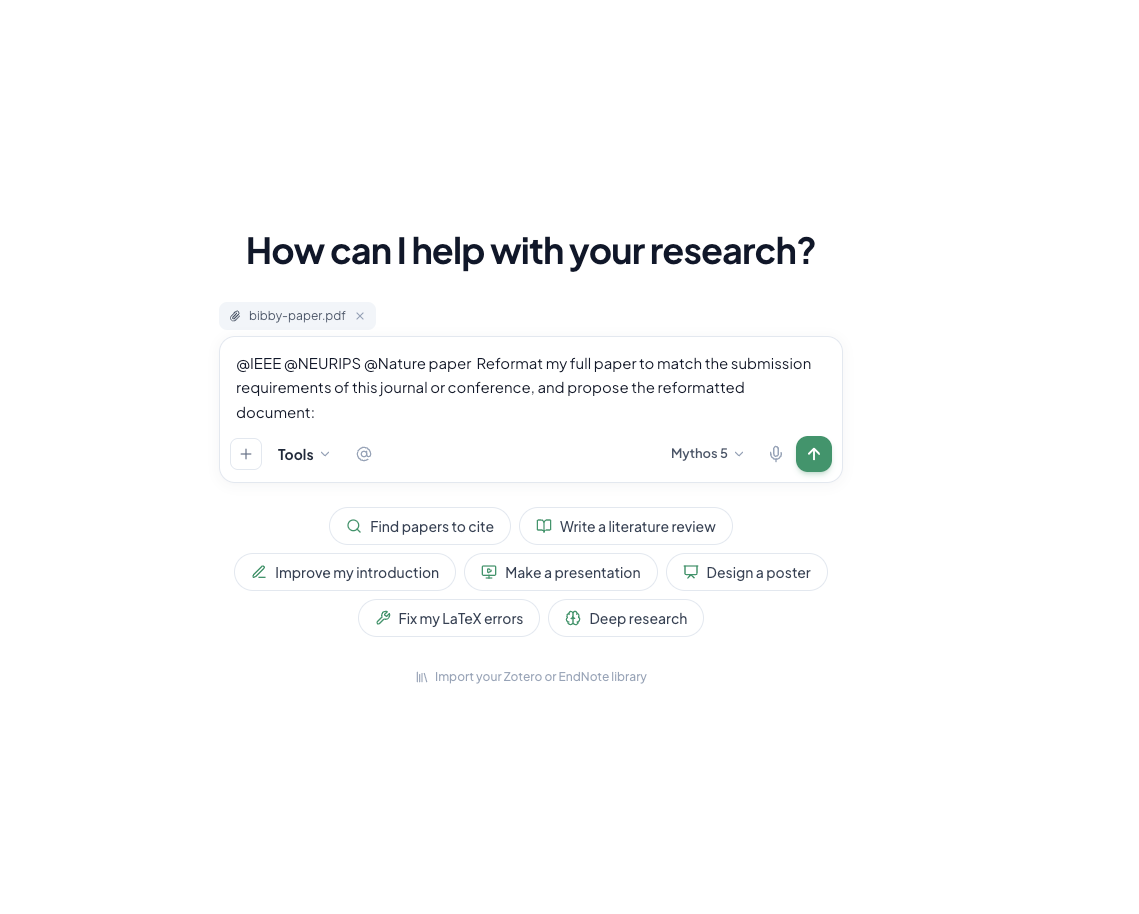}
  \caption{The \bibby{} environment. Left: source editor with structural navigation and rich-text mode toggle. Right: live compiled preview with export pipeline. The Bibby~AI panel (top right) hosts the agent layer of \cref{sec:agents}; in-document guidance (e.g., citation-search tips) is rendered inline in the compiled output during onboarding projects.}
  \label{fig:editor}
\end{figure}

Compilation is the platform's universal validator: any agent-proposed structural edit is applied to a shadow copy of the project, compiled, and rejected or repaired before the diff is offered to the author (\cref{fig:editor}).
This converts the LLM failure mode of ``plausible but broken LaTeX'' from a user-facing error into an internal retry loop.

\subsection{Ingestion Pipelines}\label{sec:ingestion}
Researchers rarely start from a blank \texttt{main.tex}.
\bibby{} provides three ingestion paths, each terminating in a compiled, editable project:
\begin{itemize}
  \item \textbf{PDF~$\to$~LaTeX}: layout-aware reconstruction of sectioning, mathematics, tables, and references from published or preprint PDFs, used for revision, extension, and template migration of existing work.
  \item \textbf{DOCX~$\to$~LaTeX}: structure-preserving conversion for collaborators arriving from Word-based workflows, mapping styles to sectioning commands and equations to \texttt{amsmath}.
  \item \textbf{Handwriting~$\to$~LaTeX}: recognition of handwritten mathematical notation into compilable expressions, targeting the whiteboard-to-manuscript gap.
\end{itemize}
Because conversion output lands inside the platform, every artifact is compile-checked immediately, and residual conversion errors are annotated in-editor rather than discovered downstream.

\subsection{Retrieval and Citation Layer}\label{sec:retrieval}
The retrieval layer answers two questions: \emph{what should I read or cite}, and \emph{why this reference over that one}.
For the first, \bibby{} performs semantic and metadata search over open scholarly indices~\cite{kinney2023semanticscholar,priem2022openalex} and inserts selected results directly as BibTeX entries with correct keys, venue fields, and in-text \verb|\cite| placement---no export/import round trip through a reference manager.

For the second, \bibby{} computes a \emph{translational impact} signal per paper by joining the scholarly record against patent-side data: PatentsView's disambiguated USPTO corpus~\cite{patentsview} and the Marx--Fuegi dataset of front-page patent citations to scientific articles~\cite{marx2020reliance}.
A paper cited by granted patents carries evidence of downstream technological use that pure academic citation counts do not capture; prior bibliometric work established this signal's validity at corpus scale~\cite{marx2020reliance}.
Surfacing it at the moment of citation choice is, to our knowledge, unique among writing platforms, and is particularly relevant for authors writing grant applications, impact statements, and applied-research papers.

\subsection{Agent Layer}\label{sec:agents}
Agents in \bibby{} are task-scoped and operate on the document model, not on selected text alone.
Following the taxonomy that has emerged in editor-assistant systems~\cite{hou2025paperdebugger}, they fall into two execution classes:
\begin{itemize}
  \item \textbf{Single-shot agents} for low-latency operations: sentence- and paragraph-level revision, notation consistency checks, and caption or abstract drafting.
  \item \textbf{Workflow agents} that coordinate retrieval, generation, and validation: literature triage into a structured related-work map; full-document review against a venue's expectations; and template retargeting, which rewrites a manuscript's preamble, sectioning, and bibliography style for a different venue and verifies the result by compilation.
\end{itemize}
All agent output is delivered as reviewable diffs against the project; nothing is applied silently.

\section{End-to-End Workflows}\label{sec:workflows}

\Cref{tab:workflow} contrasts the fragmented baseline against the \bibby{} pipeline for four recurring researcher tasks.
The structural claim is tool-count and context-switch reduction; the corresponding time measurements are drawn from platform telemetry and are the subject of our ongoing evaluation (\cref{sec:evaluation}).

\begin{table}[t]
  \centering
  \caption{Tool-chain compression for common researcher workflows. Baseline counts assume the modal stack observed in user onboarding interviews (search engine + reference manager + desktop or cloud LaTeX editor + converter utilities).}
  \label{tab:workflow}
  \begin{tabular}{lcc}
    \toprule
    Workflow & Baseline tools & \bibby{} tools \\
    \midrule
    Literature search $\to$ cited draft paragraph & 3--4 & 1 \\
    Word-collaborator manuscript $\to$ LaTeX project & 2--3 & 1 \\
    Venue rejection $\to$ reformatted resubmission & 2--3 & 1 \\
    Handwritten derivation $\to$ compiled section & 2--3 & 1 \\
    \bottomrule
  \end{tabular}
\end{table}

As one concrete trace: a user searching ``contrastive learning for tabular data'' receives ranked results annotated with academic citation counts \emph{and} patent-citation counts; selecting three papers inserts BibTeX entries and drafts a related-work paragraph with correct \verb|\cite| keys; the compile validator confirms the project builds; total interaction happens in one interface with zero copy--paste operations.

\section{Deployment and Adoption}\label{sec:deployment}

\bibby{} runs in production on containerized infrastructure (Dockerized services behind nginx on dedicated servers) with server-side compilation isolation as described in \cref{sec:editor}.
As of mid-2026 the platform serves \textbf{5{,}000+ active researchers} and \textbf{50+ subscribing universities}, with sustained month-over-month growth in both.
Institutional adoption is a meaningful signal for this product category: universities subscribe on behalf of research groups, implying evaluation against incumbent site licenses rather than individual impulse adoption.

\section{Evaluation Framework}\label{sec:evaluation}

We evaluate \bibby{} at the level of \emph{workflows}, not model outputs, because the platform's thesis is toolchain compression.
For each workflow $w$ we define researcher time cost as
\begin{equation}\label{eq:time}
  T(w) = \sum_{i=1}^{n_w} \left( t_i^{\mathrm{task}} + t_i^{\mathrm{switch}} + t_i^{\mathrm{repair}} \right),
\end{equation}
where $n_w$ is the number of tool stages, $t^{\mathrm{switch}}$ captures context-switch and transfer overhead between stages (export, upload, copy--paste, re-orientation), and $t^{\mathrm{repair}}$ captures downstream error correction attributable to stage boundaries (e.g., conversion artifacts discovered at first compile).
Editor-nativeness attacks the second and third terms directly: with $n_w = 1$, switch cost is zero by construction, and compile-validated agent output bounds repair cost.

\subsection{Modeled Time Savings}\label{sec:savings-model}

The per-workflow saving is $S(w) = T_{\mathrm{base}}(w) - T_{\mathrm{bibby}}(w)$, and the expected monthly saving per researcher is
\begin{equation}\label{eq:monthly}
  \mathbb{E}[S_{\mathrm{month}}] = \sum_{w \in \mathcal{W}} f_w \, S(w),
\end{equation}
where $f_w$ is the monthly frequency of workflow $w$.
\Cref{tab:savings} instantiates the model with stage-time estimates from onboarding interviews and task decomposition; these are \emph{modeled estimates} pending validation against production telemetry, and the parameterization is published so any term can be independently re-estimated.

\begin{table}[t]
  \centering
  \caption{Time-cost model instantiation (minutes per workflow instance). Baseline decomposition follows \cref{eq:time} over the modal fragmented stack; frequencies $f_w$ are per active-researcher month.}
  \label{tab:savings}
  \begin{tabular}{lS[table-format=3.0]S[table-format=2.0]S[table-format=3.0]S[table-format=1.1]}
    \toprule
    Workflow $w$ & {$T_{\mathrm{base}}$} & {$T_{\mathrm{bibby}}$} & {$S(w)$} & {$f_w$} \\
    \midrule
    Search $\to$ cited paragraph        & 25  & 6  & 19  & 8.0 \\
    DOCX manuscript $\to$ LaTeX project & 90  & 8  & 82  & 1.0 \\
    Venue retargeting (reformat)        & 180 & 20 & 160 & 0.5 \\
    Handwritten math $\to$ section      & 30  & 4  & 26  & 2.0 \\
    Compile-error debugging session     & 20  & 5  & 15  & 6.0 \\
    \bottomrule
  \end{tabular}
\end{table}

\begin{figure}[t]
  \centering
  \begin{tikzpicture}
    \begin{axis}[
      width=0.48\linewidth, height=5.4cm,
      ybar, bar width=9pt,
      ylabel={$S(w)$ (min saved / instance)},
      symbolic x coords={Cite,DOCX,Venue,Handwr.,Debug},
      xtick=data, x tick label style={font=\scriptsize, rotate=30, anchor=east},
      ymin=0, ymajorgrids,
      title={\small (a) Saving per workflow instance},
    ]
      \addplot coordinates {(Cite,19) (DOCX,82) (Venue,160) (Handwr.,26) (Debug,15)};
    \end{axis}
  \end{tikzpicture}\hfill
  \begin{tikzpicture}
    \begin{axis}[
      width=0.48\linewidth, height=5.4cm,
      ybar, bar width=9pt,
      ylabel={$f_w S(w)$ (min / month)},
      symbolic x coords={Cite,DOCX,Venue,Handwr.,Debug},
      xtick=data, x tick label style={font=\scriptsize, rotate=30, anchor=east},
      ymin=0, ymajorgrids,
      title={\small (b) Monthly contribution ($\Sigma = 456$ min)},
    ]
      \addplot coordinates {(Cite,152) (DOCX,82) (Venue,80) (Handwr.,52) (Debug,90)};
    \end{axis}
  \end{tikzpicture}
  \caption{Modeled time savings under the parameterization of \cref{tab:savings}. (a) Per-instance saving is largest for venue retargeting and format conversion, where the fragmented baseline is dominated by repair cost. (b) Weighted by frequency, high-recurrence workflows (citation insertion, debugging) contribute comparably to rare high-cost ones.}
  \label{fig:savings}
\end{figure}
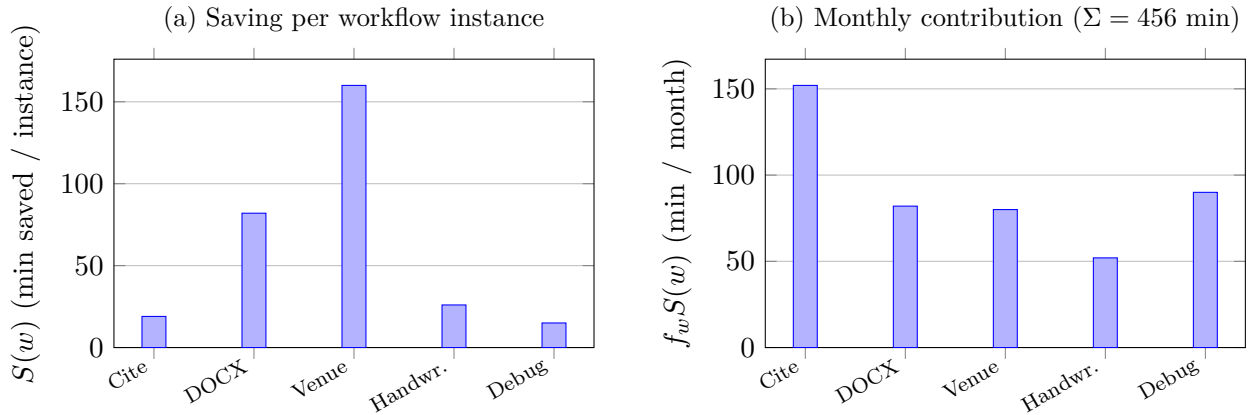

Under this parameterization, \cref{eq:monthly} yields
\begin{equation}\label{eq:result}
  \mathbb{E}[S_{\mathrm{month}}] \approx 19(8) + 82(1) + 160(0.5) + 26(2) + 15(6) = 456 \text{ min} \approx \SI{7.6}{\hour}
\end{equation}
per researcher per month---roughly \textbf{a full working day returned to research every month}, dominated by the elimination of switch and repair terms rather than by faster typing.
Aggregated across the active user base, the model implies on the order of $5{,}000 \times 7.6 \approx \num{38000}$ researcher-hours recovered monthly.
\Cref{fig:savings} visualizes the per-workflow savings and the monthly contribution of each workflow.

Our measurement program instruments the $t^{\mathrm{switch}}$ and $t^{\mathrm{repair}}$ terms in production telemetry and compares against baseline timings collected in structured onboarding interviews; the modeled values above define the pre-registered hypotheses for that measurement.

\section{Discussion and Limitations}\label{sec:discussion}

\paragraph{Owning the platform is a bet.}
Editor-nativeness trades the plugin approach's distribution advantage for architectural control.
Our adoption data suggests the trade is viable at institutional scale: a substantial share of incoming users migrate from Overleaf and comparable editors, and qualitative feedback from these switchers consistently cites the integrated pipeline as the reason for the move.
Migration friction remains the primary growth constraint, which is exactly why ingestion pipelines (\cref{sec:ingestion}) are core infrastructure rather than conveniences.

\paragraph{Patent signals are a lower bound.}
Front-page patent citations undercount science--technology linkage (in-text patent citations and non-patented use are invisible), and coverage is USPTO-centric~\cite{marx2020reliance}.
We treat the signal as evidence of impact where present, never as evidence of absence.

\paragraph{Agent trust.}
All agent edits are diff-reviewed and compile-validated, but retrieval-grounded drafting can still select real-but-suboptimal references.
Keeping the author in the loop at citation-choice time, with impact signals visible, is our current mitigation; automated claim--citation entailment checking is planned work.

\section{Conclusion}\label{sec:conclusion}

\bibby{} demonstrates that the recurring engineering obstacles of LLM-assisted academic writing---editor synchronization, unverifiable edits, fragmented retrieval---are artifacts of building assistants \emph{outside} the editor.
By making the editor, compiler, reference graph, and agents one system, the platform delivers compile-verified agentic editing, single-interface research-to-publication workflows, and a translational-impact citation signal unavailable elsewhere.
The platform is validated by production adoption across 5{,}000+ researchers and 50+ universities, and our workflow cost model estimates roughly \SI{7.6}{\hour} returned to each researcher per month---approximately \num{38000} researcher-hours across the user base---pending confirmation by the instrumented telemetry program of \cref{sec:savings-model}.

\printbibliography

\end{document}